**This preprint has now been published with smaller modifications by: Physics and Imaging in Radiation Oncology (phiRO).**

It should be cited as:

Kadhim M, Rogowski V, Persson E, Gonzalez C, Haraldsson A, Ceberg S, et al. Deep learning-based self-supervised detection and localization of magnetic resonance imaging anomalies in artificial intelligence-supported radiotherapy workflows. Phys Imaging Radiat Oncol. 2026;42
https://doi.org/10.1016/j.phro.2026.101114

It is found at:

https://doi.org/10.1016/j.phro.2026.101114

Original Research Article

# Deep learning-based self-supervised detection and localization of magnetic resonance imaging anomalies in artificial intelligence-supported radiotherapy workflows☆

Mustafa Kadhim [a,b,*], Viktor Rogowski [a,b], Emilia Persson [b,c], Camila Gonzalez [d,e], André Haraldsson [a,b], Sofie Ceberg [a], Mikael Nilsson [f], Malin Kügele [b,g], Sven Bäck [a,b], Christian Jamtheim Gustafsson [b,c]

[a] *Medical Radiation Physics, Lund University, Lund, Sweden*
[b] *Radiation Physics, Department of Hematology, Oncology, and Radiation Physics, Skåne University Hospital, Lund, Sweden*
[c] *Medical Radiation Physics, Department of Translational Medicine, Lund University, Malmö, Sweden*
[d] *Medical University of Vienna Department of Anesthesia, Intensive Care Medicine, and Pain Therapy, Austria*
[e] *Comprehensive Center for Artificial Intelligence in Medicine (CAIM), Sweden*
[f] *Centre for Mathematical Sciences, Lund University, Lund, Sweden*
[g] *Department of Radiooncology, Rostock University Medical Center, Rostock, Germany*

A R T I C L E I N F O

A B S T R A C T

**Background and purpose:** Artificial intelligence is increasingly integrated into radiotherapy workflows, yet it remains vulnerable to out-of-distribution image data that can introduce unexpected behavior in clinical tasks. Deep learning (DL)-based anomaly detection for pelvic magnetic resonance imaging (MRI) remains limited and transparent evaluation for full automation is lacking. This study aimed to develop two self-supervised anomaly-detection framework variations in radiotherapy for pelvic and brain MRI.

**Materials and methods:** A two-stage self-supervised DL-framework was trained separately for pelvic and brain MRI utilizing three public datasets. The first stage compressed normal image data into discrete representations, and a second stage modeled the representation distribution. Anomaly-evidence was obtained by combining perceptual image differences with token-surprisal scores via negative log-likelihood estimation.

Automated detection was evaluated for two anatomical cohorts: pelvis, encompassing both synthetic global anomalies and real clinical anomalies, and brain, comprising clinically annotated anomalies. Performance was assessed using sensitivity, specificity, and area under the receiver operating characteristic curve (AUC). For each cohort, a subset of normal images was reserved for analysis.

**Results:** Promising exploratory performance across evaluation cohorts was obtained. AUCs were 0.97 (95% CI 0.96–0.99) for synthetic pelvic anomalies, 0.95 (95% CI 0.90–1.00) for clinical pelvic anomalies, and 0.81 (95% CI 0.74–0.87) for brain anomalies. Qualitative heatmap analysis showed spatial agreement between detected anomalies and reference annotations.

**Conclusions:** The framework variations showed promising anomaly detection performance across pelvic and brain MRI, supporting its potential for automated MRI quality control in radiotherapy. Heatmap-based visualization further improved interpretability by spatially localizing detected anomalies.

## 1. Introduction

Artificial intelligence (AI) is rapidly being integrated into radiotherapy workflows to automate tasks such as synthetic image generation, segmentation, treatment planning, and clinical decision support [1–5]. By improving efficiency, reproducibility, and throughput while reducing manual workload, the use of AI tools has the potential to streamline clinical care [6,7]. However, safe implementation of such

☆ This article is part of a Special issue entitled: 'Physics highlights from ESTRO 2026' published in Physics and Imaging in Radiation Oncology.

* Corresponding author at: Klinikgatan 5, 222 42, Lund, Sweden.
*E-mail address:* Mustafa.kadhim@med.lu.se (M. Kadhim).

tools depends on robust image-based quality control (QC) that remains reliable not only under routine conditions, but also when exposed to unexpected inputs [8–10]. A limitation of deep learning (DL) models is their sensitivity to data distribution shifts, whereby images that differ from the training data can lead to degraded or unreliable results [11–14]. Such out-of-distribution (OOD), or anomalous, cases may arise from artefacts in acquisition, anatomical variations, pathological findings, protocol deviations, or scanner differences. If left undetected, they may produce unreliable outputs and increase the risk of clinical errors [15–17]. Robust and automated image-level anomaly detection can therefore provide quality control for automated radiotherapy workflows.

DL-based anomaly detection in magnetic resonance (MR) imaging has emerged as an active research field, predominantly focused on brain imaging [18–22]. Many of these approaches used unsupervised or self-supervised, normal-data-only training to learn a reference representation of healthy anatomy and then identify anomalies as deviations from the learned distribution. Clinically, this approach is advantageous as it eliminates the need for anomaly-specific training data and enables detection of both known and unknown anomalies.

Despite previous efforts, important knowledge gaps remain. First, prior work examined OOD detection in prostate MRI [17] and anomaly detection in female pelvic MRI [23]. Evaluation of image-level QC for prostate radiotherapy across both acquisition artefacts and clinical anomalies remains limited. Moreover, most anomaly detection publications have focused on brain MR images, leaving prostate cancer radiotherapy largely unexplored. This gap is clinically relevant as prostate radiotherapy is currently undergoing two major shifts: an increasing reliance on pelvic MR imaging (such as MRI-only workflows and MR-guided treatments) and a transition toward DL-driven care processes, with automated image reconstruction, segmentation, and treatment planning solutions already deployed in clinical settings [24–30]. These workflows rely on the high soft-tissue contrast and image quality of MR imaging for accurate target and organ-of-interest delineation, rendering the absence of image-based anomaly detection a critical limitation. Fig. 1 provides an illustrative example involving a hip implant in an MRI-only synthetic CT (sCT) workflow, motivating image-level QC.

Second, available anomaly detection frameworks were rarely assessed in the context of radiotherapy, where both local (region-specific) and global (whole-image) anomalies need to be reliably detected. In radiotherapy, multiple interdependent tasks are performed on the same voxels within the image (sCT generation, delineation, etc.), each relying on the accuracy of the previous steps. Consequently, anomalies need detection at the imaging stage. If undetected, errors can accumulate and propagate through subsequent clinical steps, introducing severe dosimetric uncertainties and compromising treatment delivery [15,31].

Third, limited attention was directed toward assessing the false-positive burden and low specificity, despite their importance in automated anomaly detection. Methods with high sensitivity at the expense of low specificity may generate false alarms, increase the clinical workload, and undermine the reliability and practical value of fully automated anomaly detection.

To address these critical vulnerabilities, this study aimed to develop and evaluate an automated anomaly detection framework for MR-guided radiotherapy, trained exclusively on normal data (i.e., in-distribution data). The framework was trained and evaluated on MR images from prostate cancer patients using the LUND-PROBE [32] dataset supplemented by both synthetic and real clinical cases. The same approach was applied to brain MR images, using models trained on the public IXI [33], fastMRI [34], and fastMRI+ [35] datasets and evaluated on the fastMRI+ anomaly dataset [35].

## 2. Materials and methods

### 2.1. *Patient cohorts*

The LUND-PROBE dataset was selected for analysis and comprised T2-weighted (T2w) pelvic imaging data from a single 3 T GE Healthcare scanner from 467 prostate cancer patients, treated within an MRI-only radiotherapy workflow at Skåne University Hospital (Lund, Sweden). All scans followed clinical protocol, centering scans around the pubic bone.

The LUND-PROBE cohort was randomly split into unique training ($n = 384$), validation ($n = 29$), and test ($n = 54$) cases for development and evaluation. Cases with hydrogel rectal spacers ($n = 5$) were extracted from the test cohort to be included as clinical anomalies. Global MRI anomalies were synthetically generated on a copy of the anomaly-free test set, including noise, ghosting, Fourier k-space spikes, motion, and blur artefacts. The intensity of global artefacts was randomly set, and simulation details are provided in Section 9 of the Supplementary material and in the code repository (Section 2.8). Details on all included anomaly categories are provided in Table 1. Additionally, we included a clinical cohort comprising retrospective pelvic MR cases with a variety of anomalies ($n = 31$) scanned on the same MR scanner, including: prostate cancer patients with hip implants ($n = 4$ scanned with the same protocol as LUND-PROBE, and $n = 4$ with different protocols); female cervical brachytherapy patients ($n = 4$); and images with a cropped scanning field of view ($n = 14$). To expand the clinical anomalies, MRI scans of a healthy volunteer ($n = 6$; 5 with global anomalies, one normal) were also included, acquired using either the

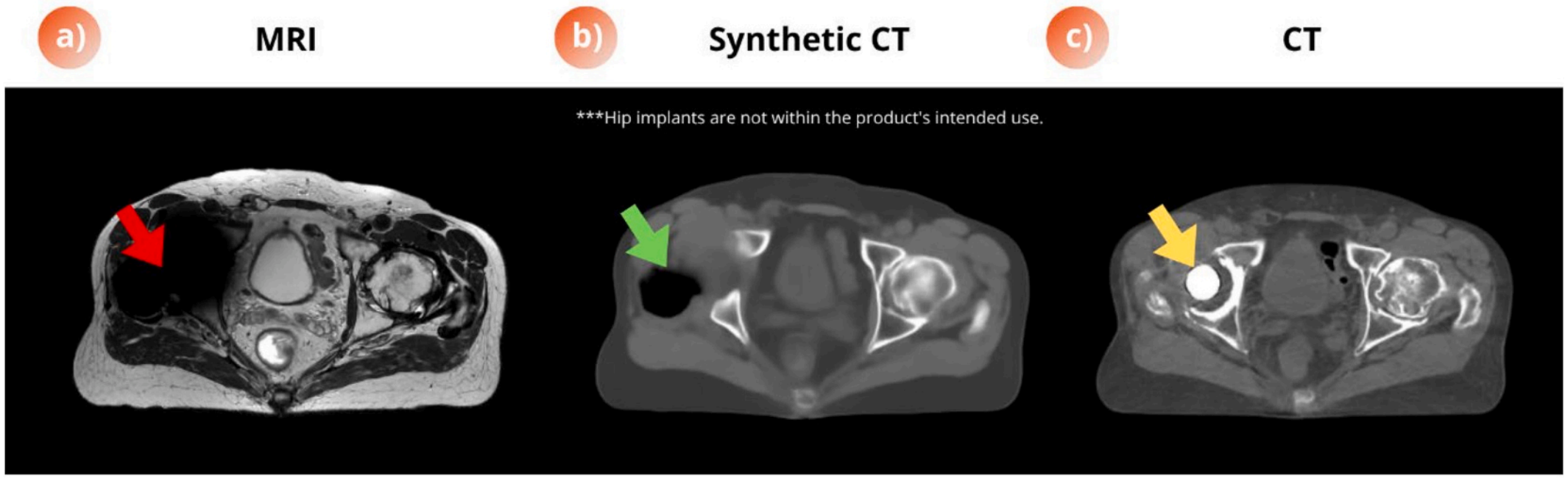

**Fig. 1.** Illustration of the need for anomaly detection in automated workflows. **(a)** A metal hip implant case in an MRI-only radiotherapy workflow, acquired using the same protocol as LUND-PROBE. The red arrow indicates a large area of signal loss caused by the metal. **(b)** A generated synthetic CT produced by a clinically approved AI-based software (MRI Planner v.2.4.14, Spectronic Medical, Helsingborg, Sweden). The green arrow highlights a hallucinated region (lack of bones and added soft-tissue). **(c)** Ground truth CT showing the accurate location of the high-density metal hip implant. The yellow arrow highlights the location of the implant. ***The use of hip implants is not within the product's intended use. This example illustrates the clinical motivation for the study and is not part of a downstream synthetic CT experimental evaluation. (For interpretation of the references to colour in this figure legend, the reader is referred to the web version of this article.)

**Table 1**
Overview of included anomaly cohorts and anomaly categories in this study.

| Cohort | Anomaly type | Global | Local |
|---|---|---|---|
| LUND-PROBE | Synthetic | Total anomalies: 242 | |
| | | Rigid motion ($n = 49$) | |
| | | Noise (n = 49) | |
| | | Ghosting (n = 49) | |
| | | Fourier space spikes ($n = 46$) * | |
| | | Blur ($n = 49$) | |
| Clinical | Real | Total anomalies: 36 | |
| | | Cropped scan field of view ($n = 14$) *** | Brachytherapy applicator ($n = 4$) *** |
| | | T1-weighted protocol (n = 1) ** | Hip implants (n = 4) *** |
| | | PD-weighted protocol (n = 1) ** | Hip implants (n = 4) |
| | | Noise ($n = 1$) ** | Hydrogel spacer ($n = 5$) |
| | | Rigid motion (n = 1) ** | |
| | | Ghosting (n = 1) ** | |
| fastMRI+ | Real | Total anomalies: 255 | |
| | | Colpocephaly (n = 1) | Extra-axial mass ($n = 6$) |
| | | Extra-axial collection ($n = 4$) | Intraventricular substance ($n = 1$) |
| | | Motion ($n = 13$) | Possible artefact ($n = 17$) |
| | | | Resection cavity ($n = 9$) |
| | | | Mass ($n = 22$) |
| | | | Craniotomy ($n = 38$) |
| | | | Posttreatment change ($n = 40$) |
| | | | Paranasal sinus opacification ($n = 2$) |
| | | | Nonspecific white matter lesion ($n = 12$) |
| | | | Absent septum pellucidum (n = 1) |
| | | | Encephalomalacia (n = 1) |
| | | | Likely cysts ($n = 2$) |
| | | | Craniectomy ($n = 1$) |
| | | | Nonspecific lesion ($n = 36$) |
| | | | Enlarged ventricles ($n = 16$) |
| | | | Dural thickening ($n = 8$) |
| | | | Lacunar infarct ($n = 4$) |
| | | | Edema ($n = 21$) |

PD: proton density, *Three patients were excluded because the randomly generated spike amplitude was too small, ** Healthy volunteer image acquisitions, *** T2-weighted protocol with image acquisition differences compared to LUND-PROBE. Visualizations of representative samples of each anomaly category are presented in Section 8 of the Supplementary material.

same protocol as LUND-PROBE or a different acquisition protocol.

For brain MRI, we used three publicly available T1-weighted (T1w) datasets: IXI, fastMRI, and fastMRI+. IXI was used solely for training ($n = 581$) and comprised brain scans from healthy participants acquired at three hospitals in London, United Kingdom. Following the fastMRI+ provided annotations (bounding box, or global labels), we identified and included both normal ($n = 172$) and anomalous scans ($n = 255$, for testing) from the fastMRI dataset. The included anomalies are listed in Table 1 and are based on the official fastMRI+ records. Normal fastMRI+ cases were randomly split into training ($n = 112$), validation ($n = 30$) and test ($n = 30$). All scans were acquired on either 1.5 T or 3 T systems with variations in acquisition protocols. The anomaly category labeled 'small-vessel chronic white-matter ischemic change' was excluded because bounding box annotations were unavailable, and lesion localization could not be reliably visually verified.

### 2.2. Ethical approval

Ethical approval for this study was provided by the regional ethics review board in Lund, diary number 2013/742, and amendment diary number 2024–01720-02.

### 2.3. Deep learning framework

A two-stage self-supervised anomaly detection framework was developed and trained exclusively on normal image data. Stage 1 learned a compact and discrete representation of normal images, whereas Stage 2 learned the distribution of normal images from the representation information provided by Stage 1. At inference, images were flagged as anomalous if both representations and perceptual appearance deviated from the learned distribution. Two separate models were trained for the pelvic and brain experiments. An overview of the framework is shown in Fig. 2; full architectural and configuration details are provided in Section 1 of the Supplementary material. Training augmentation techniques and image preprocessing details for all experiments are provided in Sections 3 and 5 of the Supplementary material.

#### 2.3.1. Stage 1: Learning discrete anatomical representation

Stage 1 used a residual vector-quantized variational autoencoder (RVQ-VAE) with a vision transformer encoder [36] and convolutional decoder to reconstruct MRI slices from discrete tokens. Two token-levels represented coarse anatomical structure (L1) and residual detail (L2) (Fig. 2a).

#### 2.3.2. Stage 2: Learning a distribution of normal reference tokens

In Stage 2, the token distribution from normal images was learned using an in-house developed factorized bidirectional transformer (Fact-biT) (Fig. 2a), inspired by the MaskGIT framework [37]. During training, a subset of tokens was randomly masked while the remaining tokens were retained as context. The model learned to iteratively predict the masked tokens from their spatial arrangement and neighboring token information. Masking and prediction were applied hierarchically: masked L1-tokens were predicted from the remaining L1 context, with L2 held fully masked, whereas L2 tokens were predicted conditioned on predicted L1.

### 2.4. Inference

Stage 1 served two functions: (1) to obtain the reconstructed image, $\hat{x}$, from the input image, x, and (2) to generate the L1 and L2 representations. An anomaly score was then obtained from a combination of two complementary pathways A and B, as shown in Fig. 2b-d.

#### 2.4.1. Pathway a: perceptual scoring

Stage 1 tokens were partially masked and restored by Fact-biT to generate a pseudo-normal ("healed") image, $\hat{x}_{Healed}$ (Fig. 2b). Learned Perceptual Image Patch Similarity (LPIPS) [38] quantified differences between the healed image and either the input or its reconstruction, as specified for each cohort. Per-pixel *Z*-scores, calibrated using normal validation slices, identified deviations from expected perceptual differences. Thresholding the Z-score map produced a binary mask, which was multiplied by the perceptual-difference map to obtain the anomaly localization map (ALM). Further thresholding produced the binary anomaly map A. Additional implementation details are provided in Section 2.2 of the Supplementary material.

#### 2.4.2. Pathway B: token-surprisal

Fact-biT predicted randomly masked L1 tokens from the remaining L1 context, with L2 fully masked (Fig. 2c). The negative log-likelihood (NLL) assigned to each observed input token quantified its unexpectedness. Values were averaged over trials in which each position was

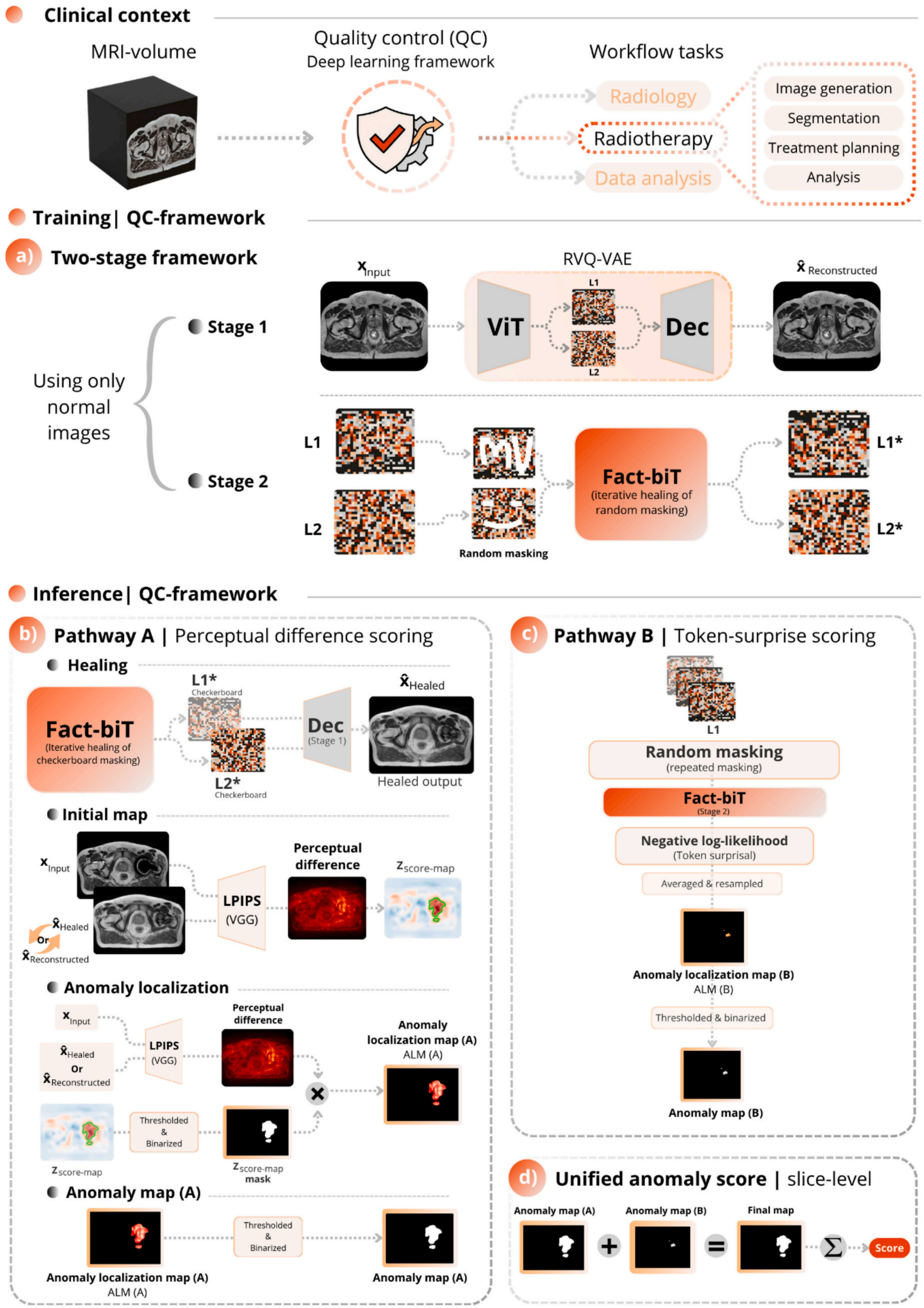

*(caption on next page)*

**Fig. 2.** Anomaly detection framework for Magnetic Resonance Imaging (MRI)-based radiotherapy quality control (QC) introduced prior to workflow tasks. Training **(a)** is self-supervised on normal image data. In Stage 1, a residual vector-quantized variational autoencoder (RVQ-VAE) with a Vision Transformer encoder (ViT) compressed each slice into two discrete token codebooks (L1,L2) and reconstructed it through a decoder (Dec). In Stage 2, a factorized bidirectional transformer (Fact-biT) learns to recover randomly masked tokens. Inference combines two complementary signals. Pathway A **(b)**, perceptual difference scoring: checkerboard-masked tokens are iteratively healed by Fact-biT and decoded into a healed image; the Learned Perceptual Image Patch Similarity (LPIPS) difference between input and healed image (or reconstructed, $\hat{x}_{Reconstructed}$, depending on task) is converted into a Z-score map ($Z_{score\text{-}map}$) using per-pixel statistics calibrated on a normal reference population. Multiplying the perceptual difference by the binarized $Z_{score\text{-}map}$ yields the anomaly localization map (ALM), later thresholded to give anomaly map A. Pathway B **(c)** token-surprisal: tokens are repeatedly masked at random, and the Fact-biT negative log-likelihood map is averaged, resized, thresholded, and binarized. High-surprisal regions, after thresholding, form anomaly map B. **(d)** The slice-level anomaly score is the total pixel count of maps A and B. Further details on the model designs and contents of Anomaly maps (A) and (B) are provided in Sections 1 and 2 in Supplementary materials.

masked, producing a token-surprisal map. Upsampling, thresholding, and binarization yielded anomaly map B. Implementation details are provided in section 2.3 of the Supplementary material.

#### *2.4.3. Unified anomaly score*

Finally, the binary anomaly maps from pathways A and B were combined and summed into a unified slice-level anomaly score (Fig. 2d). For patient-level decisions, slice scores were aggregated across all the selected slices of each scan.

### *2.5. Experimental configuration: Pelvic MRI*

Pelvic evaluation used 12 central slices (38–49) encompassing the prostate, bladder, and rectum; slice scores were summed per patient. LPIPS reference maps and the token-surprisal threshold were calibrated on 15 randomly selected validation cases. Perceptual differences were computed as LPIPS ($x$, $\hat{x}_{Healed}$) with one-sided *Z*-score thresholding. Numerical thresholds, masking settings, and test-time augmentation are detailed in Supplementary Table S2.

### *2.6. Experimental configuration: Brain MRI*

During evaluation, for each patient and anomaly category with bounding box annotations, the axial slice containing the largest annotated bounding box was selected to reduce peripheral partial-volume effects. Bounding boxes informed slice selection and qualitative heatmap assessment but were not model inputs or used to calculate anomaly scores.

LPIPS reference maps and the token-surprisal threshold were calibrated using 30 mid-axial slices from normal validation cases. To reduce scanner- and protocol-dependent noise, perceptual differences were computed between reconstructed and healed images, LPIPS ($\hat{x}$, $\hat{x}_{Healed}$), rather than LPIPS ($x$, $\hat{x}_{Healed}$) with two-sided *Z*-score thresholding. Each evaluated case received its selected slice's score. Numerical thresholds, masking settings, and test-time augmentation are detailed in Table S2 of the Supplementary material.

### *2.7. Statistical analysis*

Discrimination between anomalous and normal cases was assessed with receiver operating characteristic (ROC) analysis across the full range of anomaly-score thresholds, trading sensitivity against false-positive rate (1–specificity). Overall performance was summarized by the area under the ROC curve (AUC). We reported 95% percentile confidence intervals (CI) for AUC (2.5th to 97.5th percentiles) and a pointwise 95% ROC confidence band by evaluating stepwise true-positive rate on a fixed false-positive rate grid of 201 points. For the synthetic and clinical pelvic analyses, uncertainty was estimated using patient-level cluster bootstrapping (2000 replicates); in the clinical cohort, the six healthy volunteer acquisitions (five anomalous, one normal) formed a single cluster. Patients were sampled with replacement, retaining each selected patient's normal image and all corresponding synthetic variants together to preserve within-patient dependence. For the remaining experiments, the ROC uncertainty was estimated using a nonparametric stratified bootstrap at patient level, resampling positive and negative cases separately with replacement for 2000 replicates.

### *2.8. Code availability*

All scripts used for anomaly simulation on LUND-PROBE, cohort filtering, case selection from IXI, fastMRI, fastMRI+, training and inference of the frameworks are publicly available on GitHub [39].

## 3. Results

Results are reported separately for the pelvic and brain datasets, including ROC curves and corresponding AUC values across the full range of anomaly-score thresholds, together with qualitative anomaly heatmaps from the held-out test sets. For visualization, the anomaly localization maps ALM (A) and ALM (B) (Fig. 2b) were added, normalized per case to [0,1], and overlaid on the input image.

On the synthetic global and clinical test sets, the framework discriminated evaluated anomalies (see Table 1) from normal images with AUC of 0.97 (95% CI 0.96–0.99) and AUC of 0.95 (95% CI 0.90–1.00) for synthetic and clinical test cases (Fig. 3a). Qualitative heatmap overlays (Fig. 3b) showed spatial correspondence with the location of anomalies in both synthetic and clinical test sets. Hip implants and brachytherapy applicators were correctly detected (red arrows). For synthetic anomalies, the heatmaps were spatially diffused, consistent with anomalies that perturb wide regions of the image rather than a focal structure. From the ROC-curve, at a 5% false-positive rate, a sensitivity and specificity of 91.0% and 95.0% for synthetic and 90.9% and 95.0% for clinical sets were achieved, respectively.

On the fastMRI+ brain dataset, the framework achieved an AUC of 0.81 (95% CI 0.74–0.87; Fig. 4a). Qualitative heatmaps (Fig. 4b) showed spatial overlap with the radiologist's annotations overlaid as yellow bounding boxes. The bounding boxes were used solely to visually verify the heatmaps overlap with the annotations. Global anomalies (no bounding boxes), including motion artefacts and large extra-axial collections, produced more spatially diffused heatmaps.

## 4. Discussion

This study developed and evaluated an automated, self-supervised image-based anomaly detection framework for MRI QC in radiotherapy, trained exclusively on normal images and evaluated on publicly available, synthetic, and internal clinical datasets. To the best of our knowledge, this study was the first application of DL-based anomaly detection for MRI QC in pelvic MRI for prostate radiotherapy workflows. Furthermore, we presented the first comprehensive evaluation of the radiologist-annotated fastMRI+ and fastMRI T1w datasets in an automated anomaly detection setting. On the pelvic cohort, the framework achieved an AUC of 0.97 (95% CI 0.95–0.98) across all global synthetic artefacts and clinical test sets. In the fastMRI+ cohort, the framework achieved an AUC of 0.81 (95% CI 0.74–0.87) despite anatomical variation and differences in scanners and imaging protocols.

As radiotherapy workflows become increasingly automated, we developed a framework that could support image-level QC by flagging potentially anomalous MR images for review. However, this study did not evaluate whether anomaly scores predicted downstream errors or whether unflagged images yielded reliable downstream outputs. These relationships require validation before a clinical safety benefit can be established. Anomaly detection was also investigated at other stages of the radiotherapy workflow, including treatment-plan assessment [40], automated contour quality assurance in MR-Linac adaptive workflows [41], and, more recently, image-level anomaly detection in cone-beam CT-guided radiotherapy [42]. These studies focused on downstream radiotherapy tasks, whereas our framework centered around identifying anomalies before they enter the clinical workflow. Together, these studies and our work illustrated potential applications of anomaly detection at different stages of radiotherapy workflows.

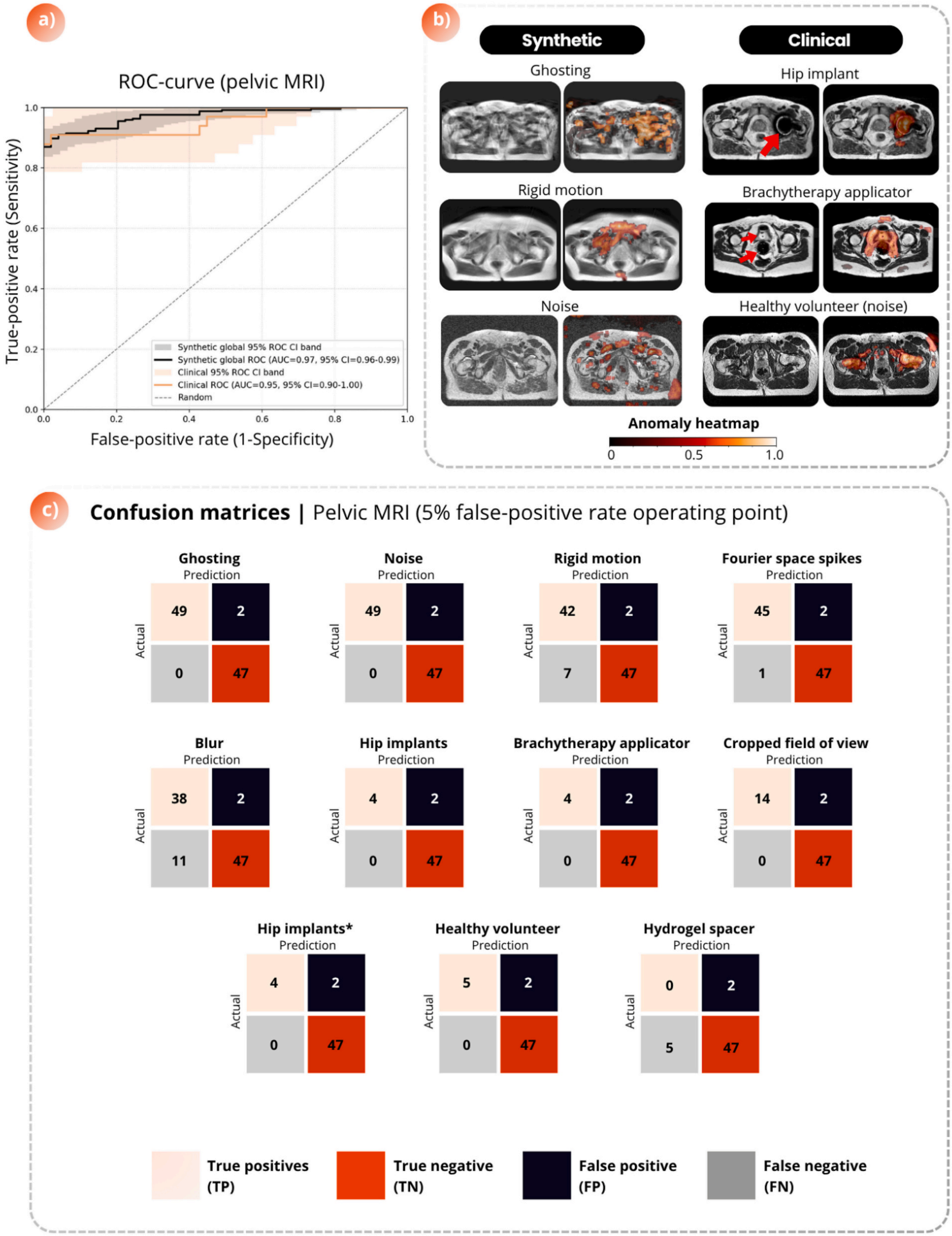

**Fig. 3.** Performance of the proposed anomaly detection framework on clinically-relevant synthetic global anomalies and clinical test set. **(a)** Patient-level receiver operating characteristic (ROC) curve showing discrimination across all possible anomaly-score thresholds for both synthetic and clinical test sets. **(b)** Qualitative anomaly heatmaps for representative synthetic global and clinical sets. Each example shows the input image (left), and the corresponding per-patient normalized anomaly heatmap overlaid on the input image (right). The heatmap represents the combination (by addition) of the perceptual and token-surprisal anomaly localization maps (ALM (A) and ALM (B)). Warmer heatmap colors denote highly anomalous findings, where the colour bar shows the normalized heatmap intensity from 0 to 1 (0 = no anomaly, 1 = maximum anomaly likelihood). Red arrows highlight the hip implant (first row) and brachytherapy applicator (second row). (c) Confusion matrices of the model performance for a selected false-positive rate of 5%. Acquisition differences relative to LUND-PROBE were present in the cropped-field-of-view, brachytherapy, and four of the eight hip-implant cases. The T1-weighted and proton-density-weighted volunteer cases represent different image contrasts. (For interpretation of the references to colour in this figure legend, the reader is referred to the web version of this article.)

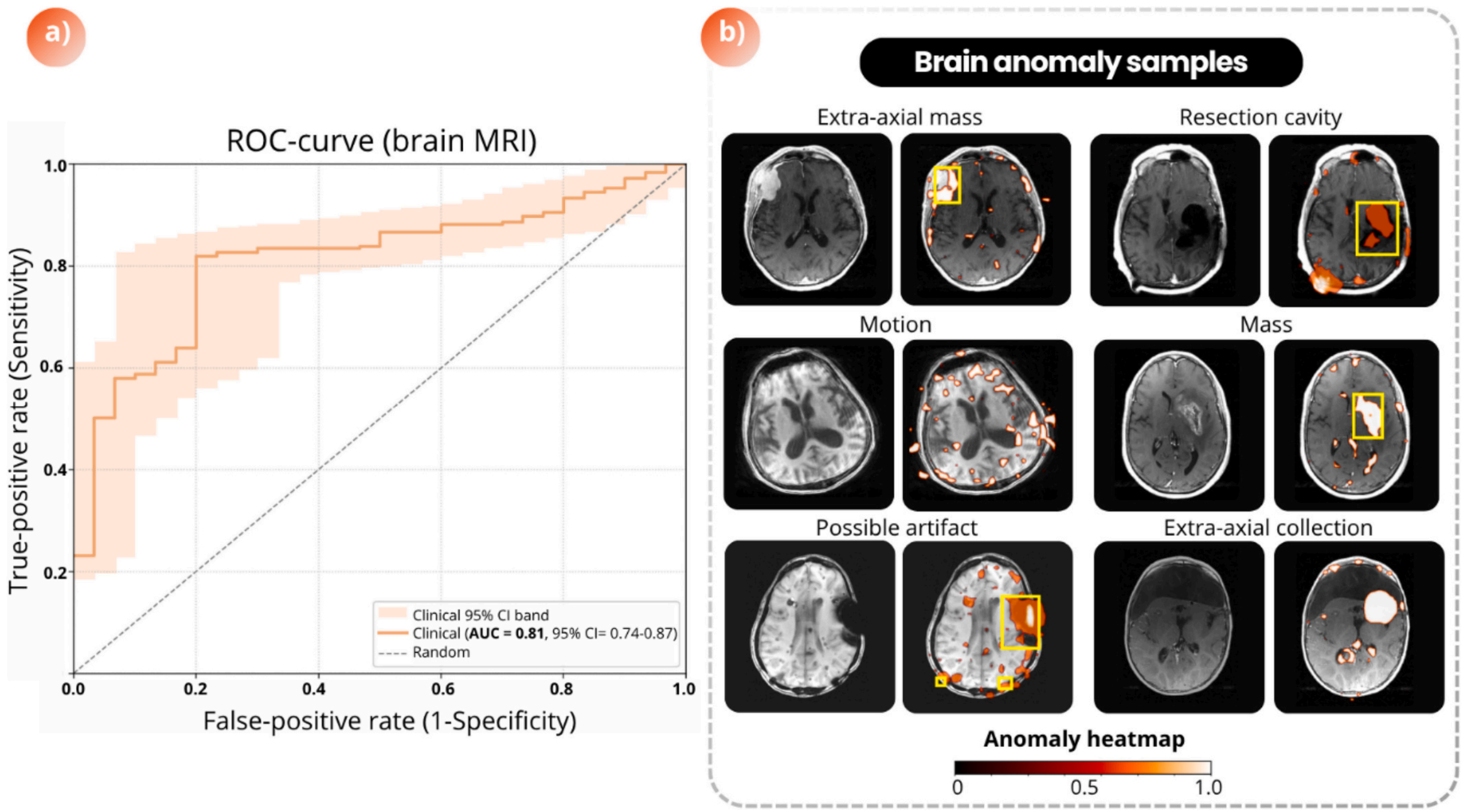

**Fig. 4.** Performance of the proposed anomaly detection framework on fastMRI+ T1w anomaly test set in Table 1. **(a)** Patient-level receiver operating characteristic (ROC) curve showing discrimination across all possible thresholds of anomaly. **(b)** Representative examples of global and local anomalies detected in fastMRI+ T1w-brain MR-images. Each example shows the input image (left), and corresponding normalized anomaly heatmap overlaid on the input image (right). The heatmap represents the combination (by addition) of the perceptual and token-surprisal anomaly localization maps (ALM-A and ALM—B). Warmer heatmap colors denote highly anomalous findings, where the colour bar shows the per-patient normalized heatmap intensity from 0 to 1 (0 = no anomaly, 1 = maximum anomaly likelihood). Yellow bounding boxes show radiologist-annotated local findings; bounding boxes were not available for global anomalies (Motion and Extra-axial collection). (For interpretation of the references to colour in this figure legend, the reader is referred to the web version of this article.)

Two design choices proved central to the framework. First, the population-based *Z*-score calibration using only normal images provided a spatial prior to expected anatomical variability. This suppressed systematic false-positive activation in regions with naturally high variability, such as that arising from rectal and bladder filling or cortical folding, so that only unusual variations were flagged. Second, the two scoring pathways (A and B) captured complementary views of normality. Token-surprisal (Pathway B) was particularly sensitive to anomalies that introduced anomalous token patterns, such as hip implants on LUND-PROBE and mass lesions on fastMRI+. Perceptual scoring (Pathway A) excelled at image-wide perturbations and diffuse findings, including motion, noise, and ghosting. Together they covered a spectrum of anomalies that neither pathway detected reliably on its own.

Several limitations merit acknowledgement. First, the anomalies introduced into the pelvis dataset were predominantly synthetic, and the clinical cohort was relatively small. Larger prospective clinical validation is therefore required before clinical use. Second, there is no universal anomaly-score threshold: site-, scanner-, and task-specific calibration is needed to set an acceptable false-positive rate. Third, performance varied across anomaly categories despite the high overall pelvic AUC. Hydrogel spacers were not detected in the five evaluated cases, potentially reflecting difficulty distinguishing their appearance from normal anatomical variability in this region. Reduced sensitivity to blurring further indicated that some image degradations were insufficiently captured by the combined anomaly score. This limitation, together with the small clinical cohort, warrants further research prior to clinical QC deployment.

Fourth, all fastMRI+ annotations were provided by a single radiologist and were limited to the axial plane, which may have introduced uncertainty into the reference labels. Fifth, the pixel-count-based score depended on the spatial extent of threshold-exceeding evidence and could therefore disadvantage small focal abnormalities. Selecting the largest annotated cross-section may have accentuated this dependence. Normalization by image area would express relative extent but would not remove the underlying size dependence. Size-stratified evaluation and alternative aggregation of anomaly evidence warrant further investigation.

Related pelvic work addressed different targets. Liu et al. [17] combined discrete latent image reconstruction with a perceptual uncertainty score derived from a pretrained prostate cancer detection network. Their evaluation linked higher uncertainty to reduced downstream cancer detection performance, whereas the present study assessed image-level anomaly discrimination without evaluating its relationship to downstream errors. Huijben et al. [43] investigated reconstruction-based detection of synthetic brain MRI artefacts and found that performance depended on the reconstruction metric and artefact type, with local artefacts more difficult to detect than global artefacts. Their findings provided context for the variation across anomaly categories observed in our study, although their pixel-based evaluation differed from our image-level analysis and datasets. Furthermore, Patsanis et al. [44] evaluated GAN-based prostate cancer detection, reporting mean internal and external test AUCs of approximately 0.71. Knupfer et al. [23] investigated disease-related anomalies in female pelvic MRI. In contrast, our evaluation concerned image-level QC for prostate radiotherapy, including synthetic artefacts and clinical anomalies. In brain MRI, Bercea et al. [19] reported that AutoDDPM detected 159 of 171 annotated abnormalities under a bounding box overlap criterion and a mid-axial slice evaluation policy; our brain AUC of 0.81 measured discrimination using a whole-image score. These results provided context, but differences in clinical targets, populations, slice selection policy, and evaluation endpoints prevented direct numerical ranking. Our framework combined perceptual differences with token surprisal to address both focal findings and image-wide artefacts, rather than relying on perceptual differences demonstrated by previous OOD methods [19,45]. Nonetheless, whether this combination improves detection relative to existing methods requires direct comparison using a common dataset and evaluation protocol.

For reliable automated QC, detection sensitivity must be balanced against unnecessary review of normal images. Our evaluation on held-out normal cases helped characterize this trade-off, although normal anatomical variability remained a potential source of false-positive findings. Calibration to the intended clinical setting is therefore important, and whole-stack brain evaluation remains necessary to establish the associated review burden. Our current brain evaluation used annotation-informed slice selection depending on the bounding box size, favoring images with a larger annotated cross-section of the abnormality. Performance may therefore be optimistic relative to unselected screening. Evaluation across complete image stacks is needed to establish examination-level sensitivity and the false-positive burden associated with reviewing normal slices.

The framework was flexible enough to accommodate diverse evaluation strategies, including heatmap-based segmentation of anomalies and hybrid schemes that combine global anomaly scores with spatial heatmap information. Future extensions may include exploring vision–language models as a second-opinion layer on flagged slices and evaluating performance for anomaly detection in MR-guided radiotherapy workflows.

In conclusion, we developed and evaluated a framework for automated detection of anomalies on MRI for radiotherapy applications trained exclusively on normal image data. Across pelvic and brain MR datasets, the framework achieved promising exploratory performance, underscoring its flexibility and potential as an automated QC tool in AI-enabled radiotherapy workflows.

## Declaration of generative AI and AI-assisted technologies in the manuscript preparation process

During the preparation of this work, the author(s) used ChatGPT and Claude for grammatical checks. The author(s) reviewed and edited the output as needed and take full responsibility for the content of the published article.

## CRediT authorship contribution statement

**Mustafa Kadhim:** Writing – review & editing, Writing – original draft, Visualization, Validation, Software, Methodology, Investigation, Formal analysis, Data curation, Conceptualization. **Viktor Rogowski:** Writing – review & editing, Visualization, Conceptualization. **Emilia Persson:** Writing – review & editing, Visualization, Supervision, Conceptualization. **Camila Gonzalez:** Writing – review & editing, Visualization, Supervision, Methodology, Conceptualization. **André Haraldsson:** Writing – review & editing, Visualization, Supervision, Conceptualization. **Sofie Ceberg:** Writing – review & editing, Visualization, Supervision, Conceptualization. **Mikael Nilsson:** Writing – review & editing, Visualization, Methodology, Conceptualization. **Malin Kügele:** Writing – review & editing, Supervision. **Sven Bäck:** Writing – review & editing, Visualization, Supervision, Conceptualization. **Christian Jamtheim Gustafsson:** Writing – review & editing, Visualization, Supervision, Resources, Project administration, Methodology, Funding acquisition, Conceptualization.

## Declaration of competing interest

The authors declare the following financial interests/personal relationships which may be considered as potential competing interests: CJG is a part time consultant for GE Healthcare. Other authors declare that they have no competing interests.

## Acknowledgements

We would like to extend our gratitude to Mizgin Coskun, MSc and Lars E Olsson, PhD Lund University, for their support on the hip implant image acquisition throughout this work.

## Appendix A. Supplementary data

Supplementary data to this article can be found online at https://doi.org/10.1016/j.phro.2026.101114.